\documentclass[10pt,prl,twocolumn,aps,hypertext]{revtex4-1}
\pdfoutput=1

\usepackage{lineno}  
\usepackage{graphicx}  
\usepackage{xspace}
\usepackage{amsmath}
\usepackage{ifthen} 
\usepackage{ifpdf}

\usepackage{hyperref}
\usepackage[all]{hypcap} 

\usepackage{amssymb}
\usepackage{amsfonts}
\usepackage{lhcb-symbols-def}

\newboolean{articletitles}
\setboolean{articletitles}{true}
\usepackage{mciteplus}

\begin{document}


\begin{titlepage}
\pagenumbering{roman}

\centerline{\Large EUROPEAN ORGANIZATION FOR NUCLEAR RESEARCH (CERN)}
\vspace*{1.0cm}
\hspace*{-0.5cm}
\begin{flushright}
  CERN-PH-EP-2011-214 \\  
  LHCb-PAPER-2011-021 \\  
  \today \\ 
\end{flushright}

\vspace*{1.0cm}

{\bf\boldmath\huge
\begin{center}
Measurement of the \CP{}-violating phase $\phijphi$\\ in the decay $\BsToJPsiPhi$
\end{center}
}

\vspace*{1.0cm}

\begin{center}
The LHCb Collaboration
\end{center}

\vspace*{5mm}

\begin{flushleft}
R.~Aaij$^{23}$, 
C.~Abellan~Beteta$^{35,n}$, 
B.~Adeva$^{36}$, 
M.~Adinolfi$^{42}$, 
C.~Adrover$^{6}$, 
A.~Affolder$^{48}$, 
Z.~Ajaltouni$^{5}$, 
J.~Albrecht$^{37}$, 
F.~Alessio$^{37}$, 
M.~Alexander$^{47}$
% ...

{\footnotesize \it
% ...
$ ^{5}$Clermont Universit\'{e}, Universit\'{e} Blaise Pascal, CNRS/IN2P3, LPC, Clermont-Ferrand, France\\
$ ^{6}$CPPM, Aix-Marseille Universit\'{e}, CNRS/IN2P3, Marseille, France\\
% ...
$ ^{23}$Nikhef National Institute for Subatomic Physics, Amsterdam, The Netherlands\\
% ...
$ ^{35}$Universitat de Barcelona, Barcelona, Spain\\
$ ^{36}$Universidad de Santiago de Compostela, Santiago de Compostela, Spain\\
$ ^{37}$European Organization for Nuclear Research (CERN), Geneva, Switzerland\\
% ...
$ ^{42}$H.H. Wills Physics Laboratory, University of Bristol, Bristol, United Kingdom\\
% ...
$ ^{47}$School of Physics and Astronomy, University of Glasgow, Glasgow, United Kingdom\\
$ ^{48}$Oliver Lodge Laboratory, University of Liverpool, Liverpool, United Kingdom\\
% ...
$ ^{n}$LIFAELS, La Salle, Universitat Ramon Llull, Barcelona, Spain\\
% ...
}
\end{flushleft}

\cleardoublepage

\begin{abstract}
  We present a measurement of the time-dependent \CP-violating
  asymmetry in \BsToJPsiPhi{} decays, using data collected with the
  LHCb detector at the LHC. The decay time distribution of
  \BsToJPsiPhi{} is characterized by the decay widths
  $\Gamma_{\mathrm{H}}$ and $\Gamma_{\mathrm{L}}$ of the heavy and
  light mass eigenstates of the \Bs-\Bsbar{} system and by a
  \CP-violating phase \phijphi{}.  In a sample of about 8500
  \BsToJPsiPhi{} events isolated from $0.37$~\invfb\ of $pp$
  collisions at $\sqrt{s}=7\TeV$ we measure $\phijphi \: = \: 0.15 \:
  \pm \: 0.18 \; \text{(stat)} \: \pm \: 0.06 \; \text{(syst) rad}$.
  We also find an average \Bs{} decay width $\Gamma_s \equiv
  (\Gamma_{\mathrm{L}}+\Gamma_{\mathrm{H}})/2 \: = \: 0.657 \: \pm \:
  0.009 \; \text{(stat)} \: \pm \: 0.008 \; \text{(syst) \invps}$ and
  a decay width difference $\DGs \equiv \Gamma_{\mathrm{L}} -
  \Gamma_{\mathrm{H}} \: = \: 0.123 \: \pm \: 0.029 \; \text{(stat)}
  \: \pm \: 0.011 \; \text{(syst) \invps}$.  Our measurement is
  insensitive to the transformation $(\phijphi,\DGs) \mapsto
  (\pi-\phijphi,-\DGs)$.

  \bigskip
  \begin{center}
    To be submitted to Physical Review Letters
  \end{center}
  
\end{abstract}

\maketitle

\end{titlepage}


\pagestyle{plain} 
\setcounter{page}{1}
\pagenumbering{arabic}

\noindent
In the Standard Model (SM) \CP{} violation arises through a single
phase in the CKM quark mixing
matrix~\cite{Kobayashi:1973fv,*Cabibbo:1963yz}.  In neutral $\B$ meson
decays to a final state which is accessible to both $\B$ and $\Bbar$
mesons, the interference between the amplitude for the direct decay
and the amplitude for decay after oscillation, leads to a
time-dependent \CP-violating asymmetry between the decay time
distributions of $\B$ and $\Bbar$ mesons.  The decay \BsToJPsiPhi{} allows
the measurement of such an asymmetry, which can be expressed in terms
of the decay width difference of the heavy (H) and light (L) \Bs{}
mass eigenstates $\DGs \equiv \Gamma_{\mathrm{L}} -
\Gamma_{\mathrm{H}}$ and a single phase
$\phi_s$~\cite{Carter:1980hr,*Carter:1980tk,*Bigi:1981qs,*Bigi:1986vr}.
In the SM, the decay width difference is $\DGs^\text{SM} = 0.087 \pm
0.021$~\invps~\cite{Lenz:2006hd,*Badin:2007bv,*Lenz:2011ti}, while the
phase is predicted to be small,
$\phi_s^\text{SM}=-2\arg\left(-V_{ts}V_{tb}^*/V_{cs}V_{cb}^*\right)=-0.036\pm0.002$~rad~\cite{Charles:2011va}.
This value ignores a possible contribution from sub-leading decay
amplitudes~\cite{Faller:2008gt}. Contributions from physics beyond the
SM could lead to much larger values of $\phi_s$~\cite{phisnewphysics}.

In this Letter we present measurements of $\phijphi$, $\DGs$ and the
average decay width $\Gamma_s \equiv
(\Gamma_{\mathrm{L}}+\Gamma_{\mathrm{H}})/2$. Previous measurements of
these quantities have been reported by the CDF and \Dzero{}
collaborations~\cite{Aaltonen:2007he,*Abazov:2008fj,*Abazov:2011ry,*Aaltonen:2011cq}.
We use an integrated luminosity of $0.37$\invfb{} of $pp$ collision
data recorded at a centre-of-mass energy $\sqrt{s}=7\TeV$ by the LHCb
experiment during the first half of 2011.  The LHCb detector is a
forward spectrometer at the Large Hadron Collider and is described in
detail in Ref.~\cite{Alves:2008zz}.

We look for \BsToJPsiPhi{} candidates in decays to $\jpsi\to\mumu$ and
$\phi\to\Kplus\Kminus$.  Events are selected by a trigger system
consisting of a hardware trigger, which selects muon or hadron
candidates with high transverse momentum with respect to the beam
direction ($\pT$), followed by a two stage software trigger.  In the
first stage a simplified event reconstruction is applied.  Events are
required to either have two well-identified muons with invariant mass
above 2.7~\gev, or at least one muon or one high-\pT{} track with a
large impact parameter to any primary vertex.  In the second stage a
full event reconstruction is performed and only events with a muon
candidate pair with invariant mass within $120$ \mev{} of the nominal
\jpsi{} mass~\cite{Nakamura:2010zzi} are retained. We adopt units
such that $c=1$ and $\hbar=1$.

For the final event selection muon candidates are required to have
$\pT > 0.5$~\gev{}. \jpsi{} candidates are created from pairs of
oppositely charged muons that have a common vertex and an invariant
mass in the range $3030-3150$~\mev{}. The latter corresponds to
about eight times the $\mumu$ invariant mass resolution and covers
part of the $\jpsi$ radiative tail. The $\phi$ selection requires two
oppositely charged particles that are identified as kaons, form a
common vertex and have an invariant mass within $\pm 12$~\mev{} of the
nominal $\phi$ mass~\cite{Nakamura:2010zzi}. The \pT{} of the $\phi$
candidate is required to exceed 1~\gev{}.  The mass window covers
approximately 90\% of the $\phi\to\Kplus\Kminus$ lineshape.

\begin{figure}[htb]
  \centerline{
    \ifthenelse{\boolean{pdf}}
      {\includegraphics[width=0.45\textwidth]{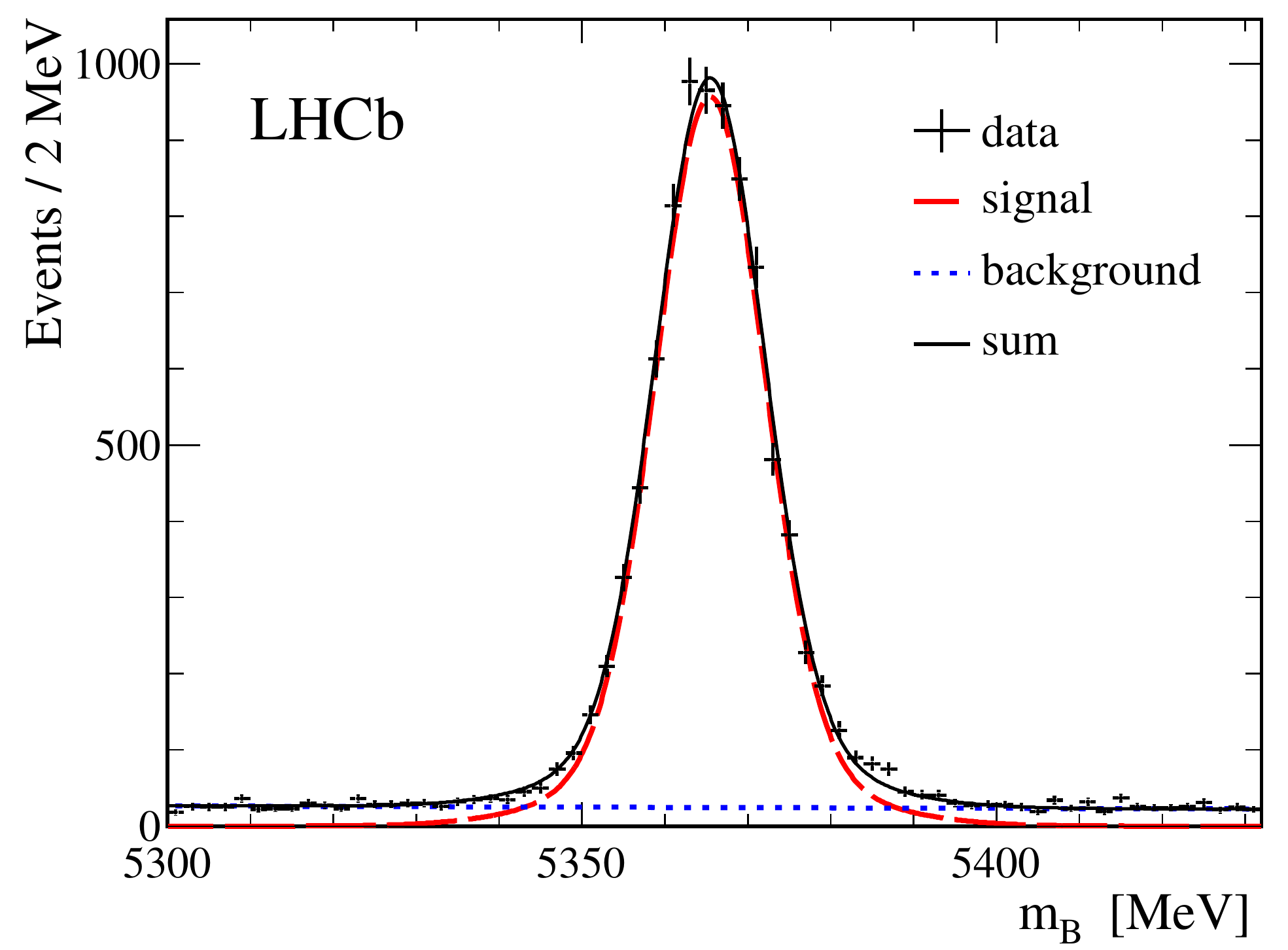}}
      {\includegraphics[width=0.45\textwidth]{figs/fig1.eps}}}
    \caption{Invariant mass distribution for $\Bs\to\mu^+\mu^- K^+K^-$
    candidates with the mass of the $\mu^+\mu^-$ pair constrained to
    the nominal $J/\psi$ mass. Curves for fitted contributions from
    signal (dashed), background (dotted) and their sum (solid) are
    overlaid.}
  \label{fig:mass}
\end{figure}

We select \Bs{} candidates from combinations of a \jpsi{} and a $\phi$
with invariant mass $m_B$ in the range $5200-5550$~\mev{}. The
latter is computed with the invariant mass of the $\mu^+\mu^-$ pair
constrained to the nominal $J/\psi$ mass.  The decay time $t$ of the
$\Bs$ is obtained from a vertex fit that constrains the
$\Bs\to\mumu\Kplus\Kminus$ candidate to originate from the primary
vertex~\cite{Hulsbergen:2005pu}.  The $\chi^2$ of the fit, which has
$7$ degrees of freedom, is required to be less than $35$. In the small
fraction of events with more than one candidate, only the candidate
with the smallest $\chi^2$ is kept.  \Bs{} candidates are required to
have a decay time within the range $0.3 < t < 14.0 \;\rm ps$.
Applying a lower bound on the decay time suppresses a large fraction
of the prompt combinatorial background whilst having a small effect on
the sensitivity to $\phijphi$.  From a fit to the $m_B$ distribution,
shown in Fig.~\ref{fig:mass}, we extract a signal of $8492 \pm 97$
events.

The $\BsToJPsiPhi{}\rightarrow \mu^+\mu^-K^+K^-$ decay proceeds via two
intermediate spin-1 particles (\emph{i.e.} with the $K^+K^-$ pair in a
P-wave).  The final state can be \CP-even or \CP-odd depending upon
the relative orbital angular momentum between the \jpsi{} and the
$\phi$.  The same final state can also be produced with $K^+K^-$ pairs
with zero relative orbital angular momentum
(S-wave)~\cite{Stone:2008ak} . This S-wave final state is \CP-odd.  In
order to measure $\phijphi$ it is necessary to disentangle the \CP-even
and \CP-odd components. This is achieved by analysing the distribution
of the reconstructed decay angles $\Omega = (\thetatr,\psitr,\phitr)$
in the transversity
basis~\cite{Dighe:1995pd,*Dighe:1998vk,Dunietz:2000cr}. In the \jpsi{}
rest frame we define a right-handed coordinate system such that the
$x$ axis is parallel to the direction of the $\phi$ momentum and the
$z$ axis is parallel to the cross-product of the \Kminus{} and
\Kplus{} momenta. In this frame $\thetatr$ and $\phitr$ are the
azimuthal and polar angles of the \mup{}.  The angle $\psitr$ is the
angle between the \Kminus{} momentum and the \jpsi{} momentum in the
rest frame of the $\phi$.

We perform an unbinned maximum likelihood fit to the invariant mass
$m_B$, the decay time $t$, and the three decay angles $\Omega$.  The
probability density function (PDF) used in the fit consists of signal
and background components which include detector resolution and
acceptance effects.  The PDFs are factorised into separate components
for the mass and for the remaining observables.

The signal $m_B$ distribution is described by two Gaussian functions
with a common mean. The mean and width of the narrow Gaussian are fit
parameters. The fraction of the second Gaussian and its width relative
to the narrow Gaussian are fixed to values obtained from simulated
events.  The $m_B$ distribution for the combinatorial background is
described by an exponential function with a slope determined by the
fit.  Possible peaking background from decays with similar final
states such as \BdToJPsiKst{} is found to be negligible from studies
using simulated events.

The distribution of the signal decay time and angles is described by a
sum of ten terms, corresponding to the four polarization amplitudes
and their interference terms. Each of these is the product of a
time-dependent function and an angular function~\cite{Dighe:1995pd,*Dighe:1998vk}
\begin{equation}
  \frac{\ud^{4} \Gamma(\BsToJPsiPhi) }{\ud t \;\ud\Omega} \; \propto \;
  \sum^{10}_{k=1} \: h_k(t) \: f_k( \Omega) \,.
  \label{Eqbsrate}
\end{equation}
The time-dependent functions $h_k(t)$ can be written as
\newcommand{\coefcosh}{\ensuremath{a_k}} \newcommand{\coefsinh}{\ensuremath{b_k}}
\newcommand{\coefcos}{\ensuremath{c_k}} \newcommand{\coefsin}{\ensuremath{d_k}}
\begin{multline}
  h_k (t) \; = \; N_k e^{- \Gs
    t} \: \left[ 
    \coefcos \cos(\dms t) \, 
    + \coefsin \sin(\dms t)
    \, \right. \\ \left. 
    + \coefcosh \cosh\left(\tfrac{1}{2} \DGs t\right) 
    + \coefsinh \sinh\left( \tfrac{1}{2} \DGs t\right)
  \right] . 
  \label{equ:timefunc}
\end{multline}
where $\dms{}$ is the \Bs{} oscillation frequency. The coefficients
$N_k$ and $a_k,\ldots,d_k$ can be expressed in terms of $\phijphi$ and
four complex transversity amplitudes $A_i$ at $t=0$. The label $i$
takes the values $\{\perp,\parallel,0\}$ for the three P-wave
amplitudes and S for the S-wave amplitude. In the fit we parameterize
each $A_i(0)$ by its magnitude squared $|A_i(0)|^2$ and its phase
$\delta_i$, and adopt the convention $\delta_0=0$ and $\sum |A_i(0)|^2 =
1$.  For a particle produced in a \Bs{} flavour eigenstate the
coefficients in Eq.~\ref{equ:timefunc} and the angular functions
$f_k(\Omega)$ are then, see~\cite{Dunietz:2000cr, Xie:2009fs}, given
by
\begin{widetext}
\newcommand{\cosphis}{\cos\phi_s}
\newcommand{\sinphis}{\sin\phi_s}
\[
\begin{array}{c|c|c|c|c|c|c}
  k  & f_k(\thetatr,\psitr, \phitr) & N_k                & \coefcosh                  & \coefsinh & \coefcos & \coefsin \\
    \hline
  1  & 2\,\cos^2\psi \left(1 - \sin^2\theta \cos^2\phi\right) & |A_0(0)|^2         & 1                          & -\cosphis & 0 & \sinphis \\
  2  & \sin^2\psi \left(1 - \sin^2\theta \sin^2\phi\right)    & |A_\|(0)|^2         & 1                          & -\cosphis & 0 & \sinphis \\
  3  & \sin^2\psi \sin^2\theta                                & |A_\perp(0)|^2      & 1                          & \cosphis & 0 & -\sinphis  \\
  4  & -\sin^2\psi \sin2\theta \sin\phi                      & |A_\|(0)A_\perp(0)| & 0                          & -\cos(\delperp-\delpar)\sinphis & \sin(\delperp-\delpar) & -\cos(\delperp-\delpar)\cosphis  \\
  5  & \tfrac{1}{2}\sqrt{2} \sin2\psi \sin^2\theta \sin2\phi & |A_0(0) A_\|(0)|   & \cos(\delpar - \delzero)   & -\cos(\delpar - \delzero) \cosphis & 0 & \cos(\delpar - \delzero) \sinphis \\
  6  & \tfrac{1}{2}\sqrt{2} \sin2\psi \sin2\theta \cos\phi   & |A_0(0) A_\perp(0)| & 0                          & -\cos(\delperp - \delzero) \sinphis  & \sin(\delperp - \delzero) & - \cos(\delperp - \delzero) \cosphis \\
  7  & \tfrac{2}{3} (1-\sin^2\theta\cos^2\phi)              & |A_\rmS(0)|^2         & 1                          & \cosphis & 0 & -\sinphis  \\
  8  & \tfrac{1}{3}\sqrt{6}\sin\psi\sin^2\theta\sin 2\phi   & |A_\rmS(0) A_\|(0)|   & 0                          & -\sin(\delpar - \delswave) \sinphis  & \cos(\delpar - \delswave) & - \sin(\delpar - \delswave) \cosphis \\
  9  & \tfrac{1}{3}\sqrt{6}\sin\psi\sin2\theta\cos\phi & |A_\rmS(0) A_\perp(0)| & \sin(\delperp - \delswave) & \sin(\delperp - \delswave) \cosphis & 0 & -\sin(\delperp - \delswave)\sinphis \\
  10 & \tfrac{4}{3}\sqrt{3}\cos\psi(1-\sin^2\theta\cos^2\phi) & |A_\rmS(0) A_0(0)|    & 0                          & -\sin(\delzero - \delswave)\sinphis & \cos(\delzero - \delswave) &  -\sin(\delzero - \delswave)\cosphis \\
\end{array}
\]
\end{widetext}

We neglect \CP{} violation in mixing and in the decay amplitudes. The
differential decay rates for a \Bsbar{} meson produced at time $t=0$
are obtained by changing the sign of $\phi_s$, $A_{\perp}(0)$ and
$A_\rmS(0)$, or, equivalently, the sign of \coefcos{} and \coefsin{} in
the expressions above.  The PDF is invariant under the transformation
$(\phijphi,\DGs,\delpar,\delperp,\delta_\rmS) \mapsto
(\pi-\phijphi,-\DGs,-\delpar,\pi-\delperp,-\delta_\rmS)$ which gives
rise to a two-fold ambiguity in the results.

We have verified that correlations between decay time and decay angles
in the background are small enough to be ignored. Using the data in
the $m_B$ sidebands, which we define as selected events with $m_B$
outside the range $5311 - 5411$~\mev{}, we determine that the
background decay time distribution can be modelled by a sum of two
exponential functions. The lifetime parameters and the relative
fraction are determined by the fit. The decay angle distribution is
modelled using a histogram obtained from the data in the $m_B$
sidebands.  The normalisation of the background with respect to the
signal is determined by the fit.

The measurement of $\phijphi$ requires knowledge of the flavour of the
\Bs{} meson at production. We exploit the following flavour specific
features of the accompanying (non-signal) $b$-hadron decay to tag the
\Bs{} flavour: the charge of a muon or an electron with large
transverse momentum produced by semileptonic decays, the charge of a
kaon from a subsequent charmed hadron decay and the momentum-weighted
charge of all tracks included in the inclusively reconstructed decay
vertex.  These signatures are combined using a neural network to
estimate a per-event mistag probability, $\omega$, which is calibrated
with data from control channels~\cite{LHCb-PAPER-2011-027}.  The
fraction of tagged events in the signal sample is
$\varepsilon_\text{tag} = (24.9 \pm 0.5)\%$.  The dilution of the \CP
asymmetry due to the mistag probability is $D = 1 - 2 \omega$.  The
effective dilution in our signal sample is $D = 0.277 \pm
0.006~\mathrm{(stat)}\pm 0.016~\mathrm{(syst)}$, resulting in an
effective tagging efficiency of $\varepsilon_\text{tag} D^2 = (1.91
\pm 0.23)\%$.  The uncertainty in $\omega$ is taken into account by
allowing calibration parameters described in
Ref.~\cite{LHCb-PAPER-2011-027} to vary in the fit with Gaussian
constraints given by their estimated uncertainties.  Both tagged and
untagged events are used in the fit.  The untagged events dominate the
sensitivity to the lifetimes and amplitudes.

To account for the decay time resolution, the PDF is convolved with a
sum of three Gaussian functions with a common mean and different
widths. Studies on simulated data have shown that selected prompt
$\jpsi\Kplus\Kminus$ combinations have nearly identical resolution to
signal events. Consequently, we determine the parameters of the
resolution model from a fit to the decay time distribution of such
prompt combinations in the data, after subtracting non-$J/\psi$ events
with the sPlot method~\cite{Pivk:2004ty} using the $\mu^+ \mu^-$
invariant mass as discriminating variable. The resulting dilution is
equivalent to that of a single Gaussian with a width of 50~fs.  The
uncertainty on the decay time resolution is estimated to be 4\% by
varying the selection of events and by comparing in the simulation the
resolutions obtained for prompt combinations and \Bs{} signal events.
This uncertainty is accounted for by scaling the widths of the three
Gaussians by a common factor of $1.00 \pm 0.04$, which is varied in
the fit subject to a Gaussian constraint. In similar fashion the
uncertainty on the mixing frequency is taken into account by varying
it within the constraint imposed by the LHCb measurement $\dms = 17.63
\pm 0.11~\mathrm{(stat)} \pm
0.02~\mathrm{(syst)}$~\invps~\cite{LHCB-PAPER-2011-010}.

The decay time distribution is affected by two acceptance
effects. First, the efficiency decreases approximately linearly with
decay time due to inefficiencies in the reconstruction of tracks far
from the central axis of the detector. This effect is parameterized as
$\epsilon(t) \propto (1 - \beta t)$ where the factor $\beta
=0.016$~\invps{} is determined from simulated events. Second, a
fraction of approximately 14\% of the events has been selected
exclusively by a trigger path that exploits large impact parameters of
the decay products, leading to a drop in efficiency at small decay
times. This effect is described by the empirical acceptance function $
\epsilon(t) \; \propto \; (a t)^c \, / \, [ 1 + (a t)^c ]$, applied
only to these events. The parameters $a$ and $c$ are determined in the
fit. As a result, the events selected with impact parameter cuts do
effectively not contribute to the measurement of \Gs{}.

The uncertainty on the reconstructed decay angles is small and is
neglected in the fit.  The decay angle acceptance is determined using
simulated events.  The deviation from a flat acceptance is due to the
LHCb forward geometry and selection requirements on the momenta of
final state particles. The acceptance varies by less than 5\% over the
full range for all three angles.

\begin{figure}[thb]
  \ifthenelse{\boolean{pdf}}{
    \includegraphics[width=0.49\columnwidth]{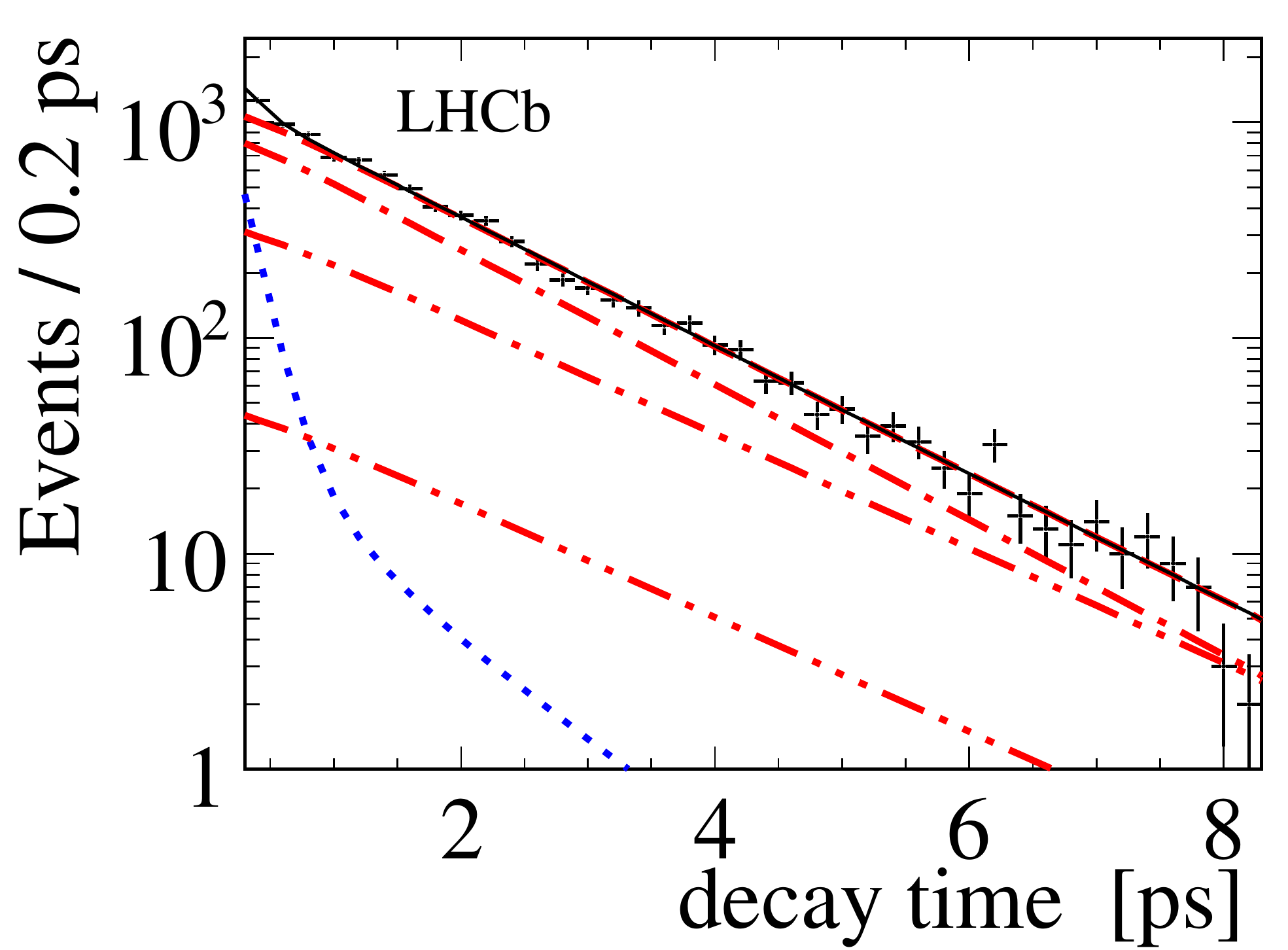}
    \includegraphics[width=0.49\columnwidth]{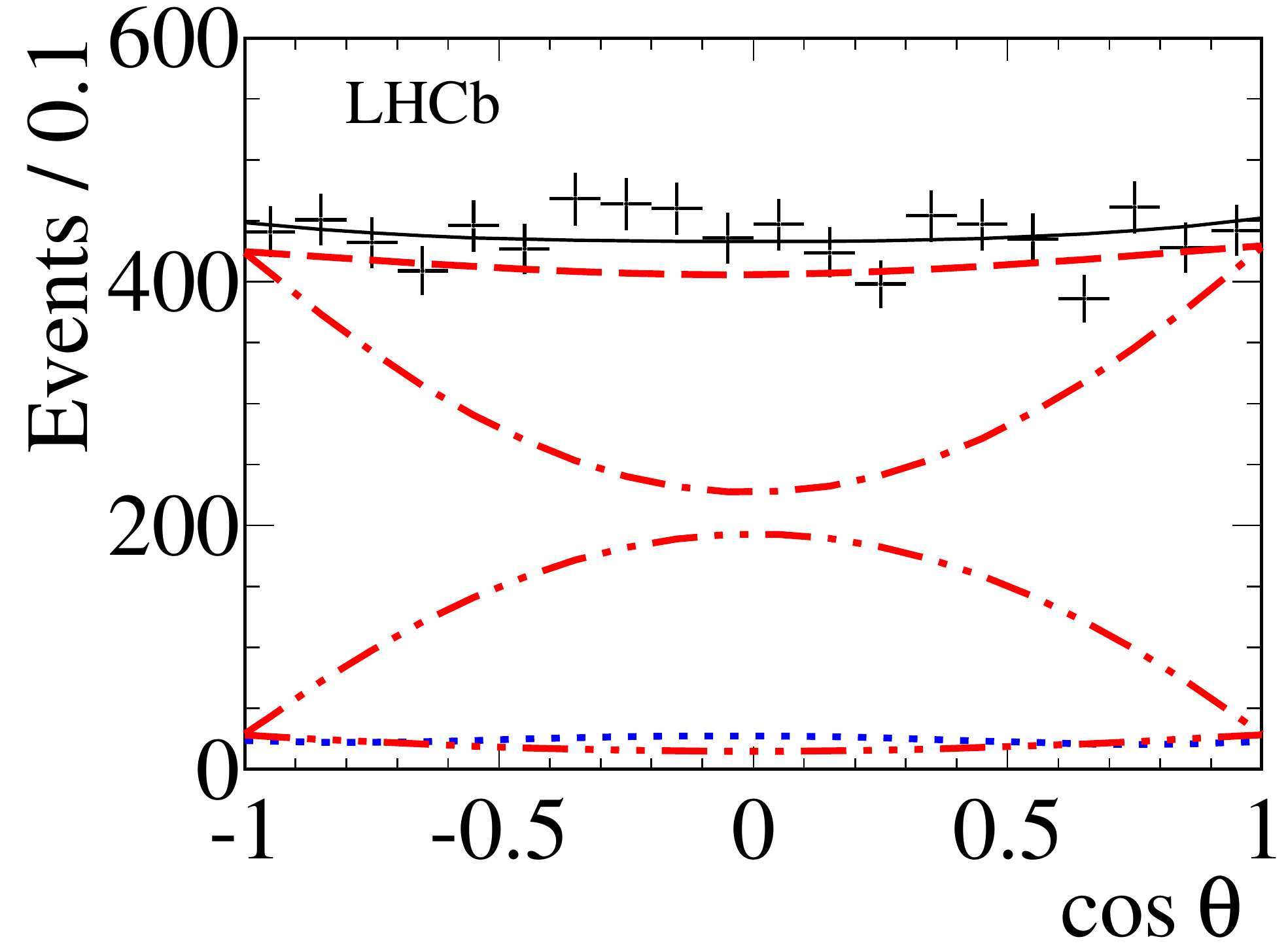}
    \includegraphics[width=0.49\columnwidth]{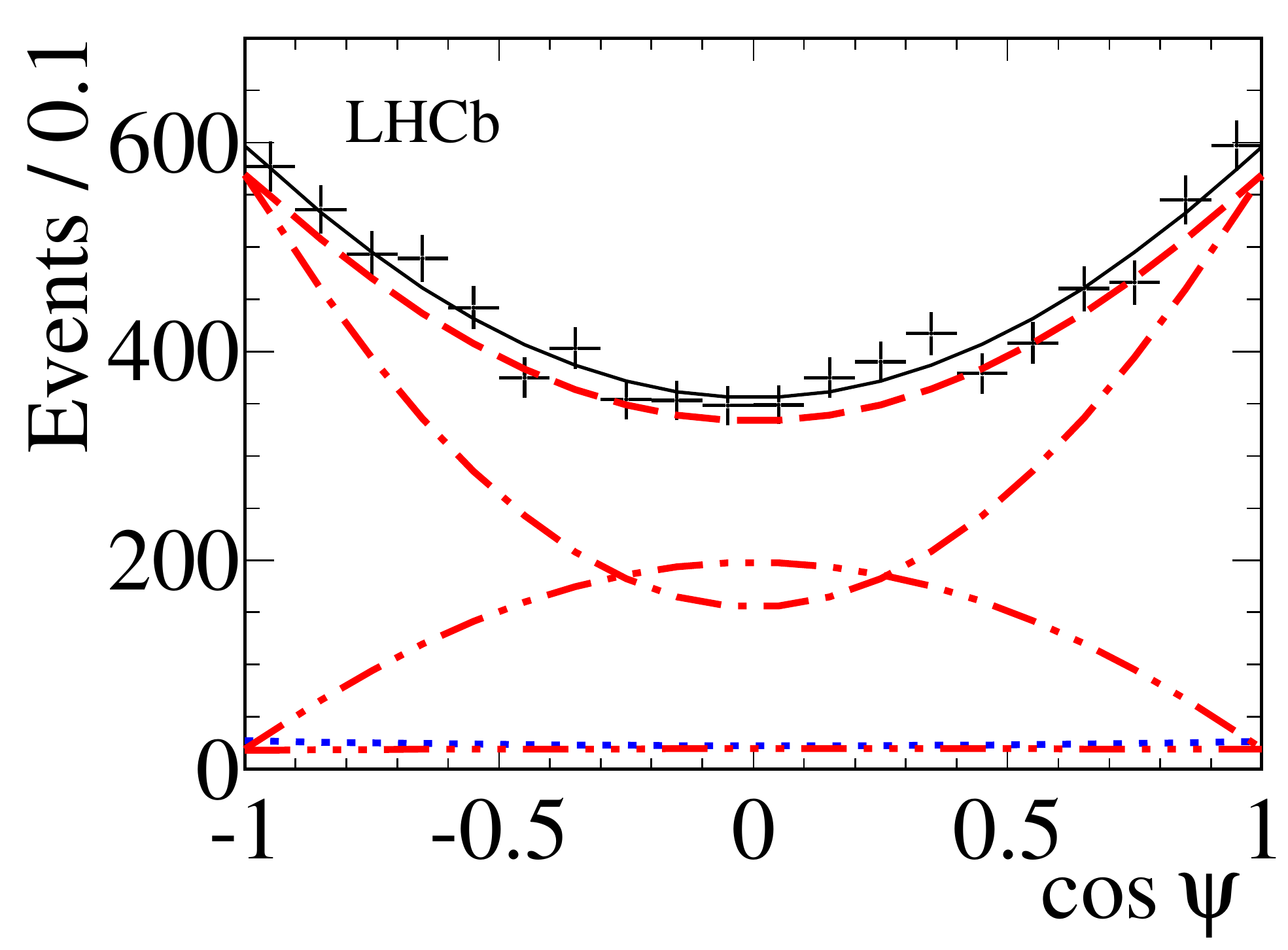}
    \includegraphics[width=0.49\columnwidth]{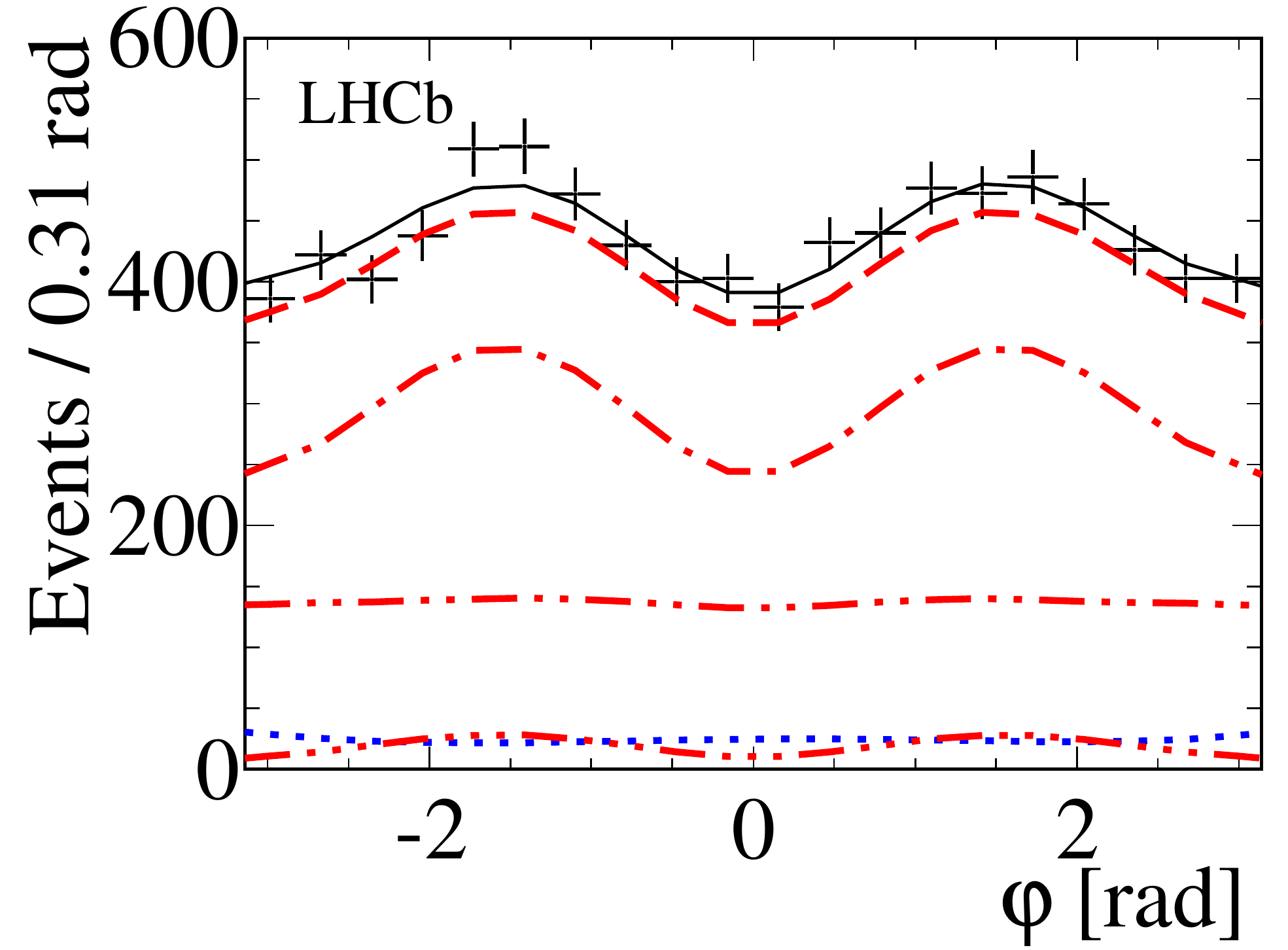}
  }{
    \includegraphics[width=0.49\columnwidth]{figs/fig2a.eps}
    \includegraphics[width=0.49\columnwidth]{figs/fig2b.eps}
    \includegraphics[width=0.49\columnwidth]{figs/fig2c.eps}
    \includegraphics[width=0.49\columnwidth]{figs/fig2d.eps}
  }
  \caption{Projections for the decay time and transversity angle
    distributions for events with $m_B$ in a $\pm\,20$ \mev{} range
    around the \Bs{} mass. The points are the data. The dashed, dotted
    and solid lines represent the fitted contributions from signal,
    background and their sum. The remaining curves correspond to
    different contributions to the signal, namely the \CP-even P-wave
    (dashed with single dot), the \CP-odd P-wave (dashed with double
    dot) and the S-wave (dashed with triple dot).}
  \label{fig:projections}
\end{figure}

\begin{figure}
  \centerline{ 
    \ifthenelse{\boolean{pdf}}
      {\includegraphics[width=\columnwidth]{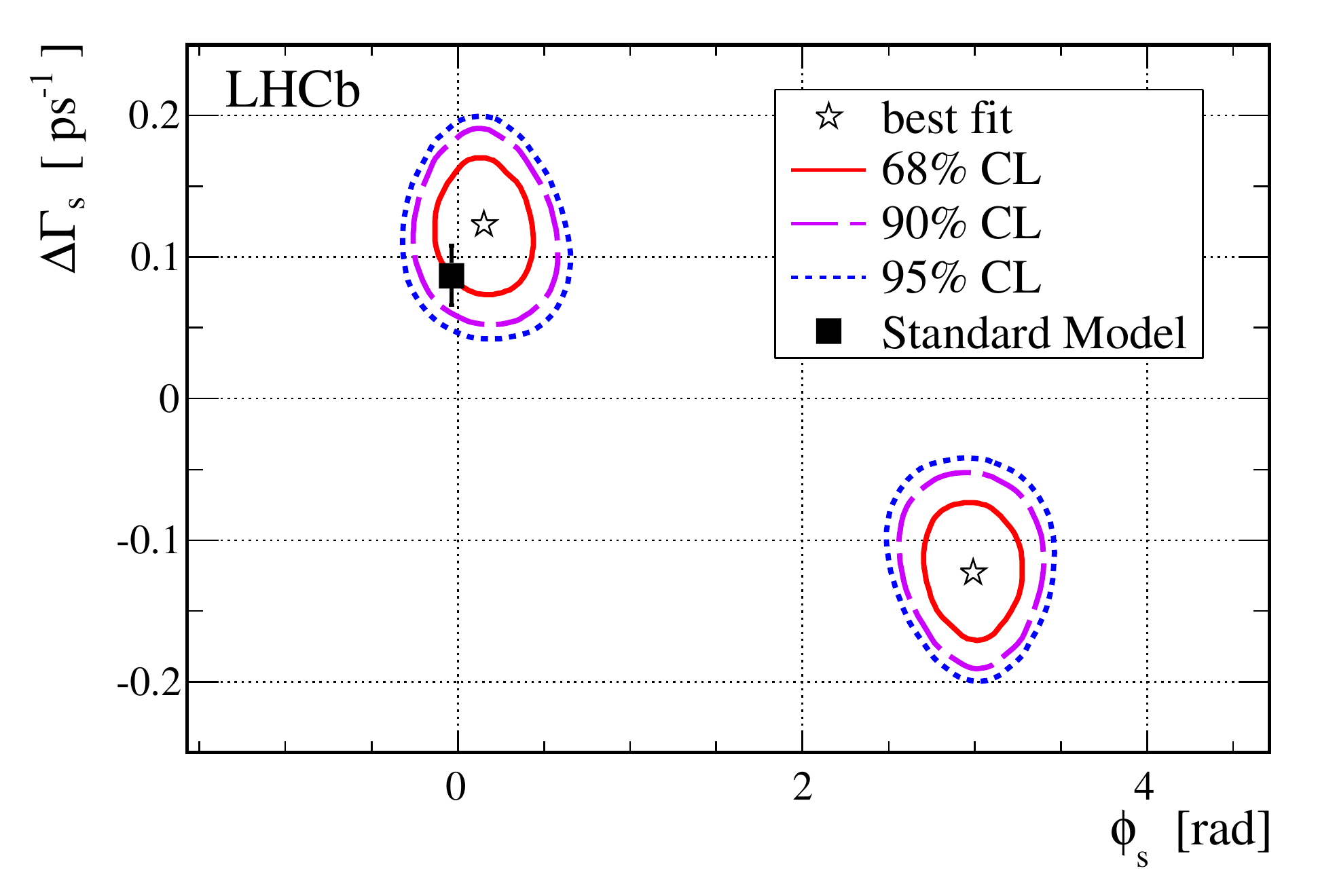}}
      {\includegraphics[width=\columnwidth]{figs/fig3.eps}}
    }
  \caption{Likelihood confidence regions in the $\DGs$-$\phijphi$
    plane. The black square and error bar corresponds to the Standard Model
    prediction~\cite{Lenz:2006hd,*Badin:2007bv,*Lenz:2011ti,Charles:2011va}.}
  \label{fig:2DLLscan}
\end{figure}

The results of the fit for the main observables are shown in Table
\ref{tab:results}.  The likelihood profile for \delpar{} is not
parabolic and we therefore quote the 68\% confidence level (CL) range
$3.0 < \delpar < 3.5$.  The correlation coefficients for the
statistical uncertainties are $\rho(\Gs,\DGs) = -0.30$,
$\rho(\Gs,\phijphi) = 0.12$ and $\rho(\DGs,\phijphi) = -0.08$.
Figure~\ref{fig:projections} shows the data distribution for decay
time and angles with the projections of the best fit PDF overlaid.  To
assess the overall agreement of the PDF with the data we calculate the
goodness of fit based on the point-to-point dissimilarity
test~\cite{Williams:2010vh}.  The $p$-value obtained is $0.68$.
Figure \ref{fig:2DLLscan} shows the 68\%, 90\% and 95\% CL contours in
the $\Delta \Gamma_s$-$\phijphi$ plane. These contours are obtained
from the likelihood profile after including systematic uncertainties,
and correspond to decreases in the natural logarithm of the
likelihood, with respect to its maximum, of 1.15, 2.30 and 3.00
respectively.

\begin{table}[ht]
  \caption{Fit results for the solution with $\DGs>0$ with statistical 
    and systematic uncertainties. \\}
  \centerline{
    \begin{tabular*}{0.75\columnwidth}{@{\extracolsep{\fill}}@{\hspace{3pt}}p{0.2\columnwidth}@{\hspace{5pt}}ccc}
      \hline
      parameter                       & value & $\sigma_\text{stat.}$ & $\sigma_\text{syst.}$\\ 
      \hline
     $\Gamma_s$ [ps$^{-1}$] \rule{0pt}{\baselineskip}    {}     & 0.657 	& 0.009 	& 0.008\\
     $\Delta \Gamma_s$ [ps$^{-1}$]   				& 0.123 	& 0.029 	& 0.011\\
     $|A_{\perp}(0)|^2$              				& 0.237 	& 0.015 	& 0.012\\
     $|A_{0}(0)|^2$                  				& 0.497 	& 0.013 	& 0.030\\
     $|A_\rmS(0)|^2$                    				& 0.042 	& 0.015 	& 0.018\\
     $\delta_{\perp}$  [rad]         				& 2.95  	& 0.37  	& 0.12\\
     $\delta_\rmS$     [rad]            				& 2.98  	& 0.36  	& 0.12\\
     \phijphi         [rad]          				& 0.15  	& 0.18  	& 0.06\\ 
      \hline
    \end{tabular*}
  }
  \label{tab:results}
\end{table}

The sensitivity to \phijphi{} stems mainly from its appearance as the
amplitude of the $\sin(\dms t)$ term in Eq.~\ref{Eqbsrate}, which is
diluted by the decay time resolution and mistag probability.
Systematic uncertainties from these sources and from the mixing
frequency are absorbed in the statistical uncertainties as explained
above. Other systematic uncertainties are determined as follows, and
added in quadrature to give the values shown in Table
\ref{tab:results}.

To test our understanding of the decay angle acceptance we compare the
rapidity and momentum distributions of the kaons and muons of selected
\Bs{} candidates in data and simulated events. Only in the kaon
momentum distribution do we observe a significant discrepancy.  We
reweight the simulated events to match the data, rederive the
acceptance corrections and assign the resulting difference in the fit
result as a systematic uncertainty. This is the dominant contribution
to the systematic uncertainty on all parameters except \Gs{}.  The
limited size of the simulated event sample leads to a small additional
uncertainty.  The systematic uncertainty due to the background decay
angle modelling was found to be negligible by comparing with a fit
where the background was removed statistically using the sPlot
method~\cite{Pivk:2004ty}.

In the fit each $|A_i(0)|^2$ is constrained to be greater than zero,
while their sum is constrained to unity. This can result in a bias if
one or more of the amplitudes is small. This is the case for the
S-wave amplitude, which is compatible with zero within $3.2$ standard
deviations. The resulting biases on the $|A_i(0)|^2$ have been determined
using simulations to be less than 0.010 and are included as systematic
uncertainties.

Finally, a systematic uncertainty of $0.008$~ps$^{-1}$ was assigned to
the measurement of \Gs{} due to the uncertainty in the decay time
acceptance parameter $\beta$.  Other systematic uncertainties, such as
those from the momentum scale and length scale of the detector, were
found to be negligible.

In summary, in a sample of $0.37$\invfb{} of $pp$ collisions at
$\sqrt{s}=7\TeV$ collected with the LHCb detector we observe $8492 \pm
97$ $\Bs\to\jpsi\Kplus\Kminus$ events with $\Kplus\Kminus$ invariant
mass within $\pm\,12$ \mev{} of the $\phi$ mass.  With these data we
perform the most precise measurements of \phijphi{}, \DGs{} and \Gs{}
in \BsToJPsiPhi{} decays, substantially improving upon previous
measurements~\cite{Aaltonen:2007he,*Abazov:2008fj,*Abazov:2011ry,*Aaltonen:2011cq}
and providing the first direct evidence for a non-zero value of
\DGs{}. Two solutions with equal likelihood are obtained, related by
the transformation $(\phijphi,\DGs) \mapsto (\pi-\phijphi,-\DGs)$. The
solution with positive \DGs{} is
\[
\setlength{\arraycolsep}{0.5mm}
\begin{array}{cclllllll}
  \phijphi & = & 0.15 & \pm &  0.18 & \text{(stat)} & \pm & 0.06 
  & \text{(syst) rad}, \\[1.5mm]
  \Gs      & = & 0.657 & \pm & 0.009 &\text{(stat)} & \pm & 0.008 
  & \text{(syst) \invps},  \\[1.5mm]
  \DGs     & = & 0.123 & \pm & 0.029 & \text{(stat)} & \pm & 0.011
  & \text{(syst) \invps},
\end{array}
\]
and is in agreement with the Standard Model
prediction~\cite{Lenz:2006hd,*Badin:2007bv,*Lenz:2011ti,Charles:2011va}.
Values of \phijphi{} in the range $0.52<\phijphi<2.62$ and
$-2.93<\phijphi<-0.21$ are excluded at 95\% confidence level. In a future
publication we shall differentiate between the two solutions by
exploiting the dependence of the phase difference between the P-wave
and S-wave contributions on the $\Kplus\Kminus$ invariant
mass~\cite{Xie:2009fs}.

\section*{Acknowledgements}

\noindent We express our gratitude to our colleagues in the CERN accelerator
departments for the excellent performance of the LHC. We thank the
technical and administrative staff at CERN and at the LHCb institutes,
and acknowledge support from the National Agencies: CAPES, CNPq,
FAPERJ and FINEP (Brazil); CERN; NSFC (China); CNRS/IN2P3 (France);
BMBF, DFG, HGF and MPG (Germany); SFI (Ireland); INFN (Italy); FOM and
NWO (The Netherlands); SCSR (Poland); ANCS (Romania); MinES of Russia and
Rosatom (Russia); MICINN, XuntaGal and GENCAT (Spain); SNSF and SER
(Switzerland); NAS Ukraine (Ukraine); STFC (United Kingdom); NSF
(USA). We also acknowledge the support received from the ERC under FP7
and the Region Auvergne.

\bibliographystyle{LHCb}
\bibliography{main}

\end{document}